\documentclass[prb,showpacs,floatfix,twocolumn]{revtex4}
\usepackage{graphics}
\usepackage{graphicx}
\usepackage[pdftitle={FeSe},
           pdfauthor={Fabio Ricci}
           ]{hyperref}
\usepackage[latin1]{inputenc}
\usepackage[dvips]{color}
\usepackage{color}
\usepackage{multirow}
\usepackage{amsmath}
\usepackage{semtrans}
\usepackage{xcolor}
\usepackage{fancyhdr}
\usepackage{epsfig}

\begin{document}

\title{van der Waals interaction in iron-chalcogenide superconductors}
\author{F. Ricci and G. Profeta}
\affiliation{CNR-SPIN and Dipartimento di Scienze Fisiche e Chimiche, Universit\`a dell'Aquila, 67010 Coppito (L'Aquila), Italy}

\begin{abstract}
We demonstrate that the inclusion of van der Waals dispersive interaction sensibly improves the prediction of lattice constants by density functional theory in
iron-chalcogenides (FeCh) superconductor compounds, namely FeSe and FeTe. We show how generalized gradient approximation (GGA) for the exchange correlation potential overestimates the out-of-plane lattice constants
in both compounds when compared with experiments. In addition, GGA predicts a too weak bonding between the neutral FeCh layers, with a sensible underestimation of the bulk
modulus. van der Waals corrected simulations completely solve both problems, reconciling theoretical results with experiments. These findings must be considered
when dealing with theoretical predictions in FeCh compounds.

\end{abstract}

\maketitle

\section{Introduction}

The discovery of high-temperature superconductivity in iron pnictides (FePn)\cite{0001,0002} raised a strong interest in searching for new 
superconducting materials which contain Fe and share common structural features with FePn superconductors. For example, superconductivity was found in the iron 
chalcogenide (FeCh) FeSe and its alloys when Se is partially substituted by Te.\cite{00003} The structure of FeCh is characterized by stack of neutral layers tetrahedrally coordinating 
Fe ions with chalcogens (Ch), similar in structure with respect to FeAs layer in FePn's which, on the contrary, are negatively charged due to the presence of intercalates
and oxides layers.\cite{review} Despite lower critical temperature, the Fe$_{1+y}$Se$_x$Te$_{1-x}$
family has a simpler crystal structure compared to those of FePn.
For these reasons, the FeCh alloys can be considered as useful prototype systems to investigate the fundamental aspects 
of structural, electronic, magnetic and superconducting properties in Fe-based superconductors.

Moreover, FeCh alloys show exceptional physical properties originating from the competing magnetic and superconducting orders.
In particular, FeSe is superconductor with a transition temperature of $T_c\sim$ 8 K at ambient pressure,\cite{00003f}
which grows up to 37 K when the pressure reaches 9 GPa,\cite{00003g} suggesting that the lattice plays a fundamental
role in the superconducting transition.
On the contrary, the parent FeTe crystal is not superconductor.\cite{00003h,00003l} However, it was shown that 90 nm thick films under tensile strain
become superconductors with onset temperature at 13 K, confirming a sensible role of the lattice.\cite{00003i}

First principles density functional theory (DFT) is considered a fundamental theoretical and computational tool to investigate the structural and electronic properties
of the normal state of Fe-based superconductors.\cite{000030l,00002} Indeed, it predicted the experimentally confirmed topology of the Fermi surfaces (FSs), the
magnetic phases and the structural distortion observed at low temperature.

In the early period of research, a strong interest was devoted to understand the predicting power of DFT in both local density (LDA) and generalized gradient approximations
(GGA),\cite{0008} showing the importance of magnetic correlation in the prediction of lattice constants. It was found that geometry optimizations
performed considering a static Fe magnetic moment (although much higher than measured one) generally mimic the magnetic fluctuations and correlations,\cite{00003m}
thus predicting crystal lattice constants in acceptable agreement with experiments.\cite{0003a}

However, a notable exception exists: {\em ab-initio} DFT in both LDA and GGA approximation fails in predicting the lattice constants
of FeCh crystal structures, in particular of the out-of-plane lattice constant.\cite{0003a} Indeed, at present, most calculations have focused on the study of electronic properties
considering the experimental measured lattice parameters with the only optimized
quantity being the Ch height (h$_{Ch}$) with respect to the Fe-atoms plane.\cite{00001,00002,00003} 
However, it must be emphasized that, due to the very simple layered structure,
the equilibrium volume and h$_{Ch}$, critically affect the electronic structure. Then, in order to predict the details of electronic properties of FeSe, FeTe
and their alloys from first principles in different physical conditions (alloys, pressure effect, surfaces etc...), it results fundamental to properly solve the theoretical problems
affecting the out-of-plane interaction between layers.

Although widely recognized,\cite{00003o} the role of dispersive van der Waals (vdW) interactions between neutral FeSe and FeTe layers
was not properly investigated.

It is well known that a general drawback of all common exchange and correlation functionals
is that they do not properly describe long-range electronic correlations,
as the vdW interaction.\cite{biork,biork2,biork3}
In fact, computational investigations using DFT are not simply interpretable when studying systems
in which vdW dispersion plays a crucial role due to non-local correlation effects.

Recently, a large effort was devoted to take into account vdW interactions and very interesting works were dedicated
to point out the state-of-the-art in advanced materials.\cite{JPhysC-s} For example, Hyldgaard, in an extensive review,\cite{JPhysC-s} evaluated and established
the limits and ranges of applicability of many different computational approaches developed to account for vdW on a large family of materials ranging from
insulators, semiconductors and metals.

In the present work, we show how first principles DFT successfully describe crystal structure of FeSe and FeTe, with unprecedented agreement
compared with experiments, once corrected to include the non-local vdW interaction, improving the calculation of out-of-plane lattice constant, interlayer binding energy and bulk modulus. In addition, we show the effect of corrected lattice
parameters in the electronic properties (band structure and Fermi surface) in both FeSe and FeTe.

\section{Computational details}

The calculations were performed using the Vienna Ab-Initio Simulation Package (VASP)\cite{vasp1,vasp2} within the generalized gradient approximation (GGA).\cite{gga}
The Perdew, Burke and Ernzerhof (PBE)\cite{00004} functional was used to calculate the exchange-correlation potential.
The GGA approximation correctly predicts the ground state magnetic phase for FeSe\cite{00001} and FeTe.\cite{00005} 
Winiarski {\em et. al.},\cite{00005a} showed that LDA estimates more precisely FeSe lattice constants than GGA, however the
calculations were performed in the PbO-type tetragonal non-magnetic phase, neglecting fundamental structural distortions and magnetic 
effects. However, the relative
success of the LDA in predicting the high temperature lattice parameters
is merely due to an accidental cancellation of errors between the correlations and exchange energies.\cite{Marini} 

In this paper, we used projected augmented-wave (PAW) pseudopotentials\cite{paw} for all the atomic species involved and in order to achieve a satisfactory
degree of convergence the 3$p^6$ 3$d^6$ 4$s^2$ states of Fe, 4$s^2$ 4$p^4$ of Se and 5$s^2$ 5$p^4$ states of Te were treated as valence electrons with an energy cutoff 
up to 550 eV. 
Integrations over the Brillouin Zone (BZ) was performed 
considering different uniform Monkhorst and Pack grids\cite{kpoints} depending on lattices: 13$\times$13$\times$9 and
14$\times$7$\times$9 for magnetic collinear stripe (AFM1) FeSe ($a\sqrt{2}\times b\sqrt{2}\times c$ unit cell),
and magnetic bicollinear double stripe (AFM2) FeTe ($a\times 2b \times c$ crystal unit cell), respectively. For the tetragonal ($a\times a\times c$) paramagnetic (PM)
phase, which contains two Fe and two Ch atoms, we used 20$\times$20$\times$15 and 15$\times$15$\times$9 k-grid for FeSe and FeTe, respectively.

The vdW interaction is considered using the DFT-D2 Grimme's semi-empirical force-field correction\cite{0004,00004q} and the so-called
vdW-optB86b functional as implemented in the VASP code.\cite{Klimes} The two functionals were chosen for their simplicity (DFT-D2) and high
accuracy (vdW-optB86b).

\begin{figure}[h!]
\begin{center}
\begin{tabular}{ccc}
\includegraphics[clip,scale=0.48]{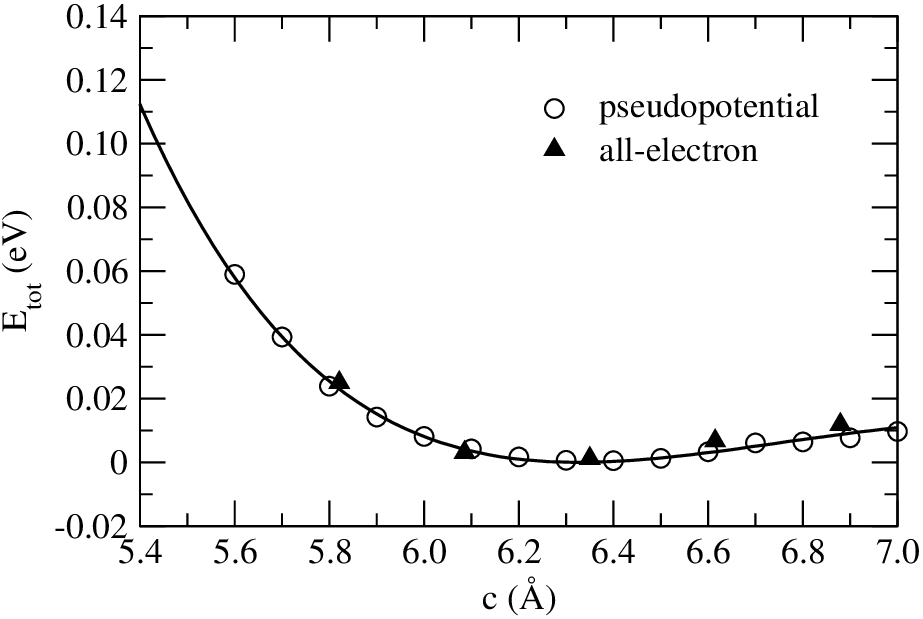}
\end{tabular}
\caption{The pseudopotential (open circles) and all-electron (full triangles) energy curves calculated for FeSe AFM1 phase as a function of $c$
lattice constant. The in-plane $a$ and $b$
lattice constants are kept fixed at the experimental values.\cite{0009} The solid black line is a guide for the eyes.}
\label{fig:0000}
\end{center}
\end{figure}

Due to the high accuracy required in the calculations, we previously checked the pseudopotential quality with all-electron full potential linear augmented plane-wave
method in the FLAIR implementation.\cite{jansen1,weinert1}
In Fig. \ref{fig:0000} we show the $c$-axis relaxation, fixing the in-plane lattice parameter
to experimental values (see below and Tab. \ref{tab:001} for all details and references) on AFM1 FeSe for VASP
and FLAIR simulations. The pseudo-potential energy curve nicely agrees with the all-electron one in a wide range of $c$
lattice constant, giving equilibrium $c$ of 6.30 and 6.25, respectively. This is a fundamental consistence check due to the already discussed issues
related to the comparison between all-electron within the well converged pseudopotentials.\cite{0008}

\section{Results and discussion}

\begin{figure}[h!]
\begin{center}
\begin{tabular}{cc}
\includegraphics[clip,scale=0.48]{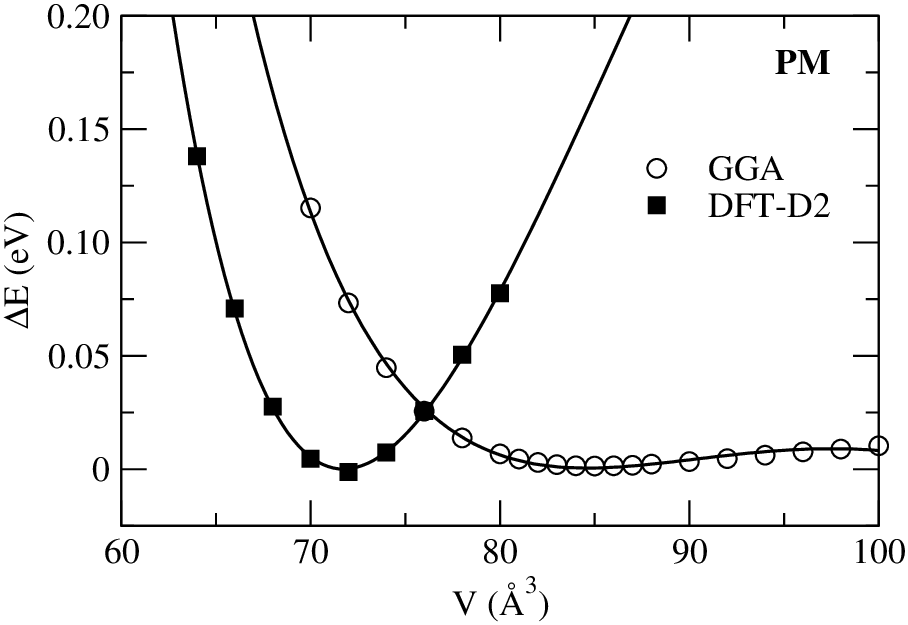}\\
\includegraphics[clip,scale=0.48]{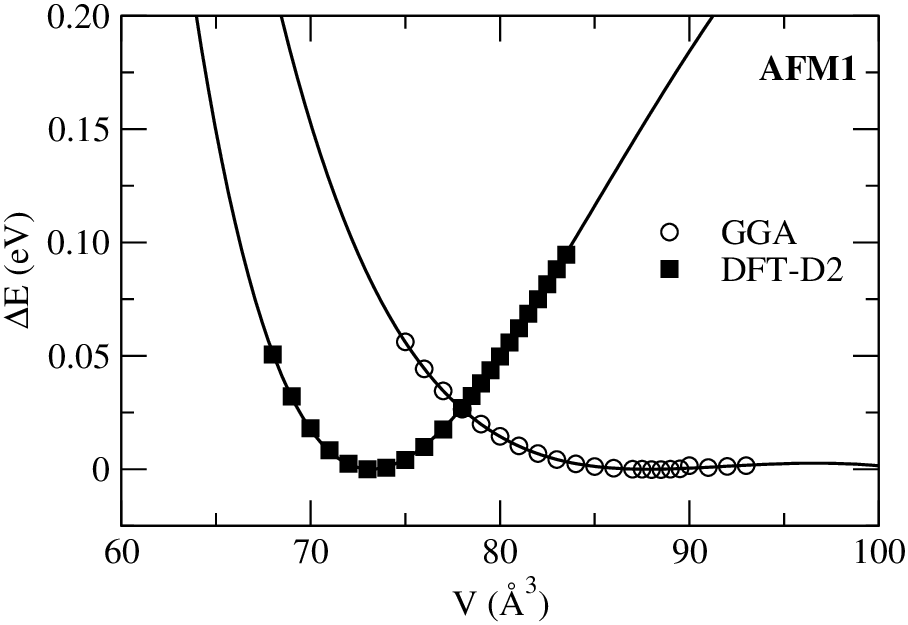}
\end{tabular}
\caption{FeSe energy-volume curves for GGA (open circles) and vdW DFT-D2 (full squares) in the PM (upper panel) and AFM1 (lower panel)
phases. Volume and energies refer to the conventional unit cell.
Solid lines are the fitted Murnaghan curve.\cite{0004a}}
\label{fig:01}
\end{center}
\end{figure}

\begin{figure}[h!]
\begin{center}
\begin{tabular}{cc}
\includegraphics[clip,scale=0.48]{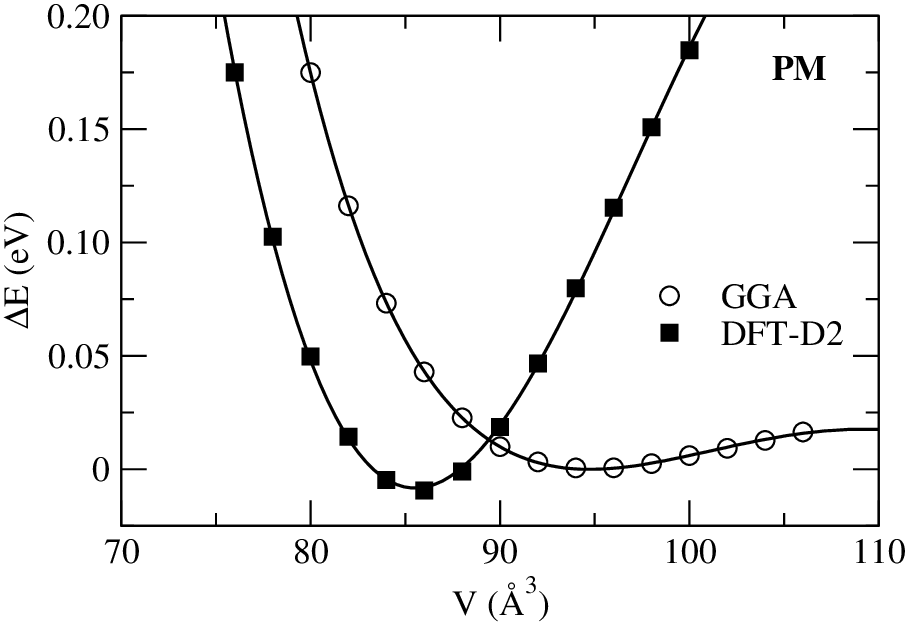}\\
\includegraphics[clip,scale=0.48]{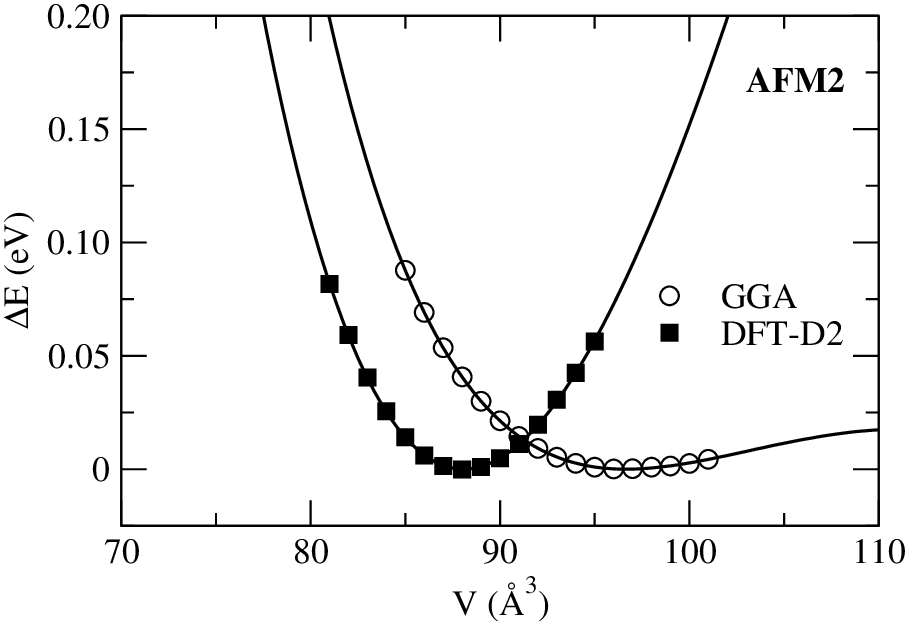}
\end{tabular}
\caption{FeTe energy-volume curves for GGA (open circles) and vdW DFT-D2 (full squares) in the PM (upper panel) and AFM2 (lower panel)
phases. Volume and energies refer to the conventional unit cell.
Solid lines are the fitted Murnaghan curve.\cite{0004a}}
\label{fig:02}
\end{center}
\end{figure}

Having tested the accuracy of pseudopotentials, we calculated the energy-volume phase diagrams for FeSe and FeTe in both PM
and AFM configurations with and without the vdW correction. The results (with available experimental measurements) are presented
in Tab. \ref{tab:001} and Fig. \ref{fig:01} and \ref{fig:02}.  

\begin{center}
\begin{table}[h!]
		\caption{Structural parameters and magnetic moments for FeSe and FeTe.
		Available experiments are also reported.}
		\vspace{0.5cm}
		\begin{center}
		\scalebox{1}{
		\begin{tabular}{|c|c|c|c|c|c|c|}
		\hline
		EXP & T (K) & $a $ (\AA) & $b $ (\AA) & $c $ (\AA) & h$_{Ch}$ (\AA)&  $\alpha$ (deg) \\
		\hline
		FeSe\cite{0009} & 7 & 3.7646 & 3.7540 & 5.479 & 1.4621  & 90.00 \\
		FeTe\cite{0005} & 2 & 3.8312 & 3.7830 & 6.264 & 1.7540  & 89.17 \\
		\hline
		\hline
		GGA & {} & $a $ (\AA) & $b $ (\AA) & $c $ (\AA) & h$_{Ch}$ (\AA) & $\alpha$ (deg) \\
		\hline
		FeSe&\multirow{2}{*}{PM} & 3.68 & 3.68 & 6.26 & 1.39 & 90.00 \\
		FeTe && 3.81 & 3.81 & 6.52 & 1.59 & 90.00 \\
		\hline
		FeSe&\multirow{2}{*}{AFM} & 3.75 & 3.71 & 6.32 & 1.45  & 90.00 \\
		FeTe && 3.87 & 3.63 & 6.90 & 1.78 & 86.60 \\
		\hline
		\hline
		DFT-D2 &{} & $a $ (\AA) & $b $ (\AA) & $c $ (\AA) & h$_{Ch}$ (\AA) & $\alpha$ (deg) \\
		\hline
		FeSe&\multirow{2}{*}{PM} & 3.64 & 3.64 & 5.42 & 1.40  & 90.00  \\
		FeTe && 3.77 & 3.77 & 6.03 & 1.59  & 90.00  \\
		\hline
		FeSe&\multirow{2}{*}{AFM} & 3.67 & 3.61 & 5.53 & 1.46  & 90.00 \\
		FeTe & & 3.81 & 3.61 & 6.42 & 1.77 & 88.53 \\
		\hline
		\end{tabular}}
		\end{center}
		\label{tab:001}
\end{table}
\end{center}

Since the PM calculations neglect magnetic interactions, fundamental to reproduce experiments,
we will mainly focus our attention on AFM phases.

In both FeSe and FeTe the GGA curves show a very weak interaction between the layers,
predicting a too large lattice $c$ constant when compared with experiments.
As we can see, for FeSe (FeTe) the GGA gives a $\sim$15\% (10\%) deviation from experiments for the out-of-plane $c$ lattice constant, while the
in-plane $a$ and $b$ parameters are in good agreement with a deviation lower than 1\% (4\%).

On the contrary, the interlayer bonding energy, corrected with the semi-empirical DFT-D2 vdW dispersion potential, shifts the
minimum of the total energy indicating an increased interaction between FeCh layers, shrinking
all the lattice parameters with respect to
GGA. In FeSe and FeTe $a$ and $b$ remain consistent with experimental values in a range of $\sim$0\%-5\%, depending
on material, while along the $c$-axis, there is a sensible improving: the 
theoretical value is corrected within 1-3\% with respect to experiments. Moreover, it is very interesting to note that, for FeTe AFM phase, the DFT-D2 improve also the
monoclinic $\alpha$ angle. 

We have fitted the curves with a Birch-Murnaghan state equation,\cite{0004a} and compare the equilibrium volumes and the bulk moduli (shown in Tab. \ref{tab:003})
with available experiments.

\begin{center}
\begin{table}[h!]
		\caption{Equilibrium volumes and bulk moduli of the conventional cell calculated for paramagnetic (PM) and antiferromagnetic (AFM: AFM1 and AFM2 for
		FeSe and FeTe, respectively) phases with and without vdW correction.}
		\vspace{0.5cm}
		\begin{center}
		\scalebox{1}{
		\begin{tabular}{|c|c|c|c|}
		\hline
		 EXP & T (K) & $V_{eq} $ (\AA$^3$) & B$_0$ (GPa)   \\
		\hline
		FeSe\cite{bulk1} & 50 & 77.56 & 33 \\
		FeTe\cite{bulk3} & 300 & 91.98 & 36\\
		\hline
		\hline
		 GGA & {} & $V_{eq} $ (\AA$^3$) & B$_0$ (GPa)   \\
		\hline
		FeSe & \multirow{2}{*}{PM}& 84.95 & 5.28  \\  
		FeTe && 94.73 & 9.81  \\
		\hline
		FeSe & \multirow{2}{*}{AFM} & 87.82 & 3.37  \\  
		FeTe & & 96.65 & 9.71  \\
		\hline
		 DFT-D2 & {} & $V_{eq} $ (\AA$^3$) & B$_0$ (GPa)  \\
		\hline
		 FeSe &\multirow{2}{*}{PM} & 71.73 & 37.57  \\
		 FeTe & & 85.53 & 36.46  \\
		\hline
		 FeSe &\multirow{2}{*}{AFM}& 73.18 & 34.67  \\
		 FeTe & & 88.08 & 38.99   \\
		\hline
		\end{tabular}}
		\end{center}
		\label{tab:003}
\end{table}
\end{center}

The most evident result is the striking disagreement between GGA and experimental bulk moduli for both FeSe and FeTe. In fact, as evident from Fig. \ref{fig:01}
and \ref{fig:02}, the out-of-plane interaction is too weak resulting in a very low B$_0$. This behaviour is dependent on the magnetic phase considered.
Interestingly, we observe that the vdW correction completely changes the physics and chemistry of the out-of-plane interaction thus resulting in a much better
agreement with experiments. We note that, even in this case, both PM and AFM phases are corrected in the same way.
The overall satisfactory agreement indicates that the vdW interlayer interaction is fundamental to correctly
reproduce the properties of FeCh compounds. To the best of our knowledge, the bulk moduli of both FeSe and FeTe have never been predicted so far.

As evident from our results, and as already well discussed in review articles (see, for example, Ref. \onlinecite{biork,biork2,JPhysC-s}),
the main effect of the vdW interaction is the $c$ lattice constant reduction and the consequent increase of B$_0$, an effect
common to other layered materials.

In recent reviews,\cite{JPhysC-s} it was shown as different approaches to include the vdW interaction can lead to different calculated lattice constants.
The DFT-D2 is a semi-empirical method, very efficient, but relies on the optimization of four semi-empirical parameters,\cite{0004} previously fitted
on different classes of materials.\cite{00004q} In order to further investigate vdW functionals, we used the so-called vdW-optB86b,\cite{Klimes,Klimesa} which includes
the non-local vdW interaction in the exchange and correlation energy functionals. This method was demonstrated to have a wide range of applicability and excellent
agreement with experimental results on different solids in term of lattice constants, bulk moduli and atomization energies. We performed lattice parameters
and internal coordinates optimization varying independently $a$ and $c$ and compared the results with DFT-D2 method. The results
show the complete consistence of the two approaches.

The use of the corrected $c$ lattice constant has a strong effect on band structure and FS near the Fermi energy. In order to understand this effect on electronic
properties, we calculated the band structure and FSs for AFM FeSe and FeTe considering both GGA and DFT-D2 relaxed $c$ lattice constants.
To disentangle the (small) differences in the in-plane lattice parameters (see Tab. \ref{tab:001}) we calculated the equilibrium $c$ constant fixing
both $a$ and $b$ ones to experimental values. This is a well justified procedure to predict interlayer distance widely used in literature in the
case of layered crystals in which the
strongest vdW correction comes from the out-of-plane interaction.\cite{biork,biork2}
 
\begin{figure}[h!]
\begin{center}
\begin{tabular}{ccc}
\includegraphics[clip,scale=0.48]{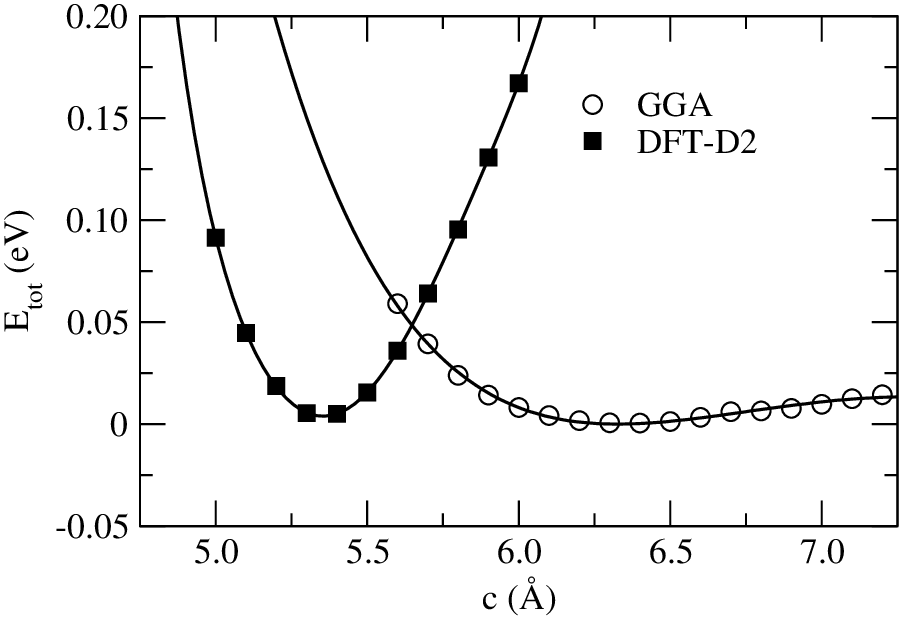}\\
\includegraphics[clip,scale=0.48]{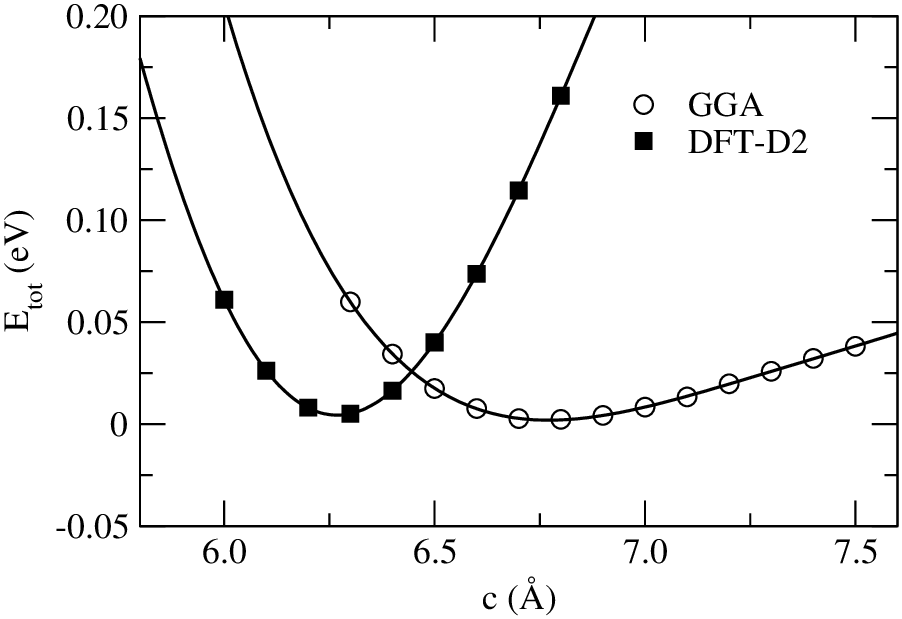}
\end{tabular}
\caption{AFM1 FeSe (upper panel) and AFM2 FeTe (lower panel) $c$-relaxed total energy GGA (open circles) and DFT-D2 (full squares) curves. The solid lines are shown as guides for the eyes.}
\label{fig:00001}
\end{center}
\end{figure}

Fig. \ref{fig:00001} show the theoretical results obtained in this way for AFM FeSe and FeTe phases
which are compatible with experiments in a range around 2\% and 0.1\% for the FeSe and FeTe, respectively.

\begin{figure}[h!]
\begin{center}
\begin{tabular}{ccc}
\includegraphics[clip,scale=0.48]{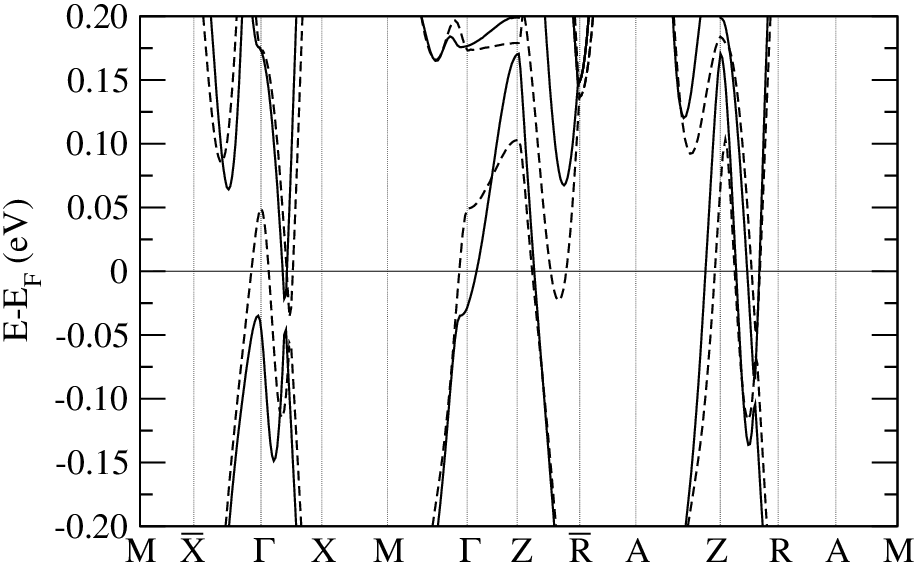}
\end{tabular}
\caption{FeSe band structure calculated using the GGA (dashed lines) and DFT-D2 (solid lines) lattice constants.
}
\label{fig:00003}
\end{center}
\end{figure}

\begin{figure}[h!]
\begin{center}
\begin{tabular}{ccc}
\includegraphics[clip,scale=0.48]{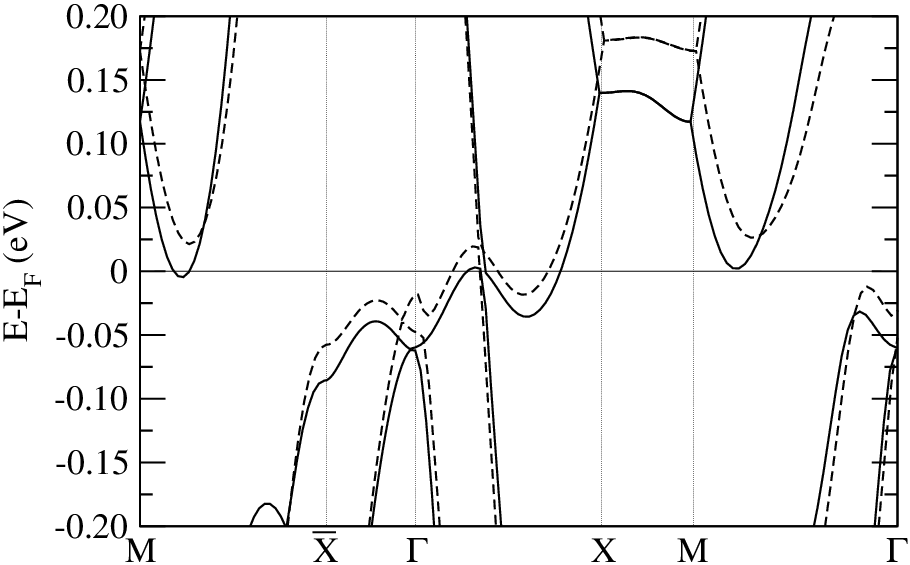}
\end{tabular}
\caption{FeTe band structure calculated using the GGA (dashed lines) and DFT-D2 (solid lines) lattice constants.
}
\label{fig:00004}
\end{center}
\end{figure}

\begin{figure}[h!]
\begin{center}
\begin{tabular}{ccc}
\includegraphics[clip,scale=0.6]{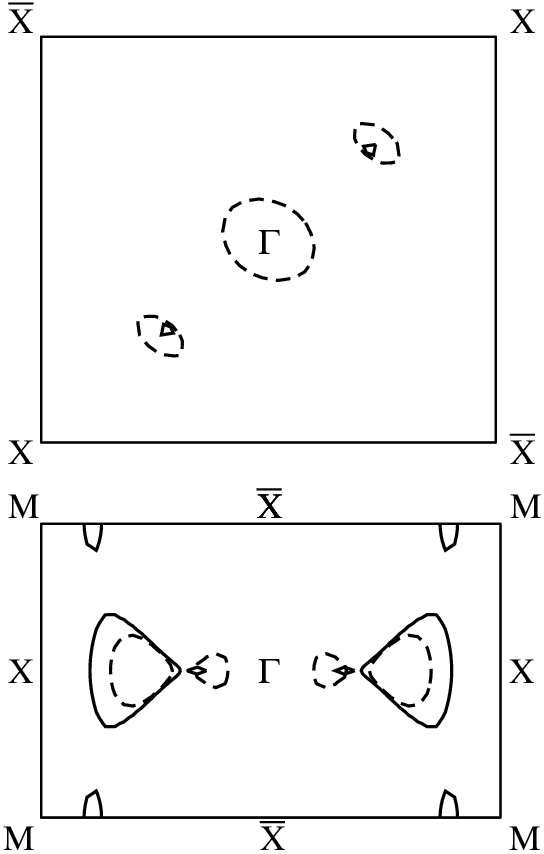}
\end{tabular}
\caption{FeSe and FeTe Fermi surfaces (upper and lower panel, respectively) calculated using GGA (dashed line) and DFT-D2 (solid line) lattice constants.}
\label{fig:00006}
\end{center}
\end{figure}

In Fig. \ref{fig:00003} and \ref{fig:00004} we show the band structures calculated for the FeSe and FeTe using the above lattice constants, and in Fig. \ref{fig:00006}
the relative FSs. Considering the GGA $c$ parameter, we observe in FeSe an hole pocket at the $\Gamma$ point and two very small electron ones along the $\Gamma$X
direction. In particular, these two pockets are related to the presence of a Dirac-like point just below the Fermi energy.
A nearly Dirac point is also present on $c^{*}/2$ plane, along the ZR line.
The electronic states changes sensibly using the predicted $c$ lattice constant with vdW correction: the hole pocket at the $\Gamma$ point is now completely filled, while
the Dirac point along the $\Gamma$X line shifts nearer E$_F$, closing all FSs in the $\Gamma$XM plane.

In FeTe we observe that hole and electron pockets along the $\Gamma$X for the GGA $c$ lattice
constant transforms in electron pocket completely filling the hole one, once vdW parameters are considered.

In conclusion, we studied the effect of the vdW correction on the calculation of lattice constants and bulk modulus of FeCh superconductors.
We showed that the vdW correction is fundamental in order to predict lattice structure and bulk moduli in agreement with the experiments, having a large effect on the out-of-plane
bonding between the Ch-Fe-Ch layers. 

These results are important in view of computational experiments within first-principles DFT methods on Fe-based superconductors and
can also be extended to predict the effect of substitutions, intercalations, high pressure, strain and surface effects on the structural, electronic and magnetic properties
of these compounds. 
\newpage

\begin{acknowledgments}
This work was supported by FP7 European project SUPER-IRON (grant agreement No. 283204).
The work was supported by a CINECA-HPC ISCRA grant and by an HPC grant at CASPUR.
\end{acknowledgments}

\end{document}